\documentclass[aps,preprint,floatfix]{revtex4}
\usepackage[dvips]{graphics,graphicx}
\usepackage{amsmath}
\begin{document}

\title{Quantum entanglement and the Born-Markov approximation \\ for an open quantum system}
\author{Andrey~R.~Kolovsky$^{1,2}$}
\affiliation{$^1$Kirensky Institute of Physics, 660036 Krasnoyarsk, Russia}
\affiliation{$^2$Siberian Federal University, 660041 Krasnoyarsk, Russia}
\date{\today}
\begin{abstract}
We revisit the Born-Markov approximation  for an open quantum system by considering a microscopic model of the bath, namely, the Bose-Hubbard chain  in the parameter region where it is chaotic in the sense of Quantum Chaos. It is shown that strong ergodic properties of the bath justify all approximations required for deriving the Markovian master equation from the first principles. 
\end{abstract}
\maketitle

%%%%%%%%%%%%%%%%%%%%%%%%%%%%%%%%%%%%%%%%%%%%%%
{\em 1.} Nowadays one witnesses a recovery of interest to open quantum systems, with the emphasis shifted to open many-body systems \cite{Witt08,Barm11,Bran12,115,Labo16,113,Kord15,108}. The standard approach to dynamics of an open system is the master equation for the reduced density matrix,
%***********************************
\begin{equation}
\label{1}
\rho_S(t)={\rm Tr}_B[ R(t)] \;,
\end{equation}
where $R(t)$ is the total density operator of the combined system consisting of the system of interest (subindex $S$) and the bath (subindex $B$). In the Markovian case dynamics of the matrix  $\rho_S(t)$ is governed by the Lindblad equation \cite{Davi76,Carm91,Breu07,wiki},
%************************************
\begin{equation}
\label{2}
\frac{{\rm d} \rho_S}{{\rm d}t}=-i[H_S,\rho_S]  
%\end{equation}
%\begin{displaymath}
-\sum_n \frac{\gamma_n}{2} \left(\rho_S V_n^\dag V_n - 2 V_n \rho_S V_n^\dag + V_n^\dag V_n \rho_S \right)  \;,
%\end{displaymath}
\end{equation}
where $H_S$ is the system Hamiltonian and $\gamma_n$ and $V_n$ are the (problem specific)  relaxation constants and operators. Mathematically, this structure of the master equation is fixed by the condition of positivity for the density matrix \cite{Lind76}. However, the microscopic derivation of Eq.~(\ref{2}) is a tedious procedure  which involves a number of approximations. These approximations are usually summarised  as follows: (i) Interaction between the system and the bath is weak; (ii) Correlation time for the relevant bath observables  is much shorter than the characteristic time-scale of the system dynamics; (iii) At any time the total density matrix factorises into the tensor product of the reduced density matrices, 
%***************************************
\begin{equation}
\label{3}
R(t) = \rho_B(t) \otimes \rho_S(t) \;.
\end{equation}
We mention that that the last equation, which is explicitly or implicitly used in every tutorial on the master equation,  never holds. 
%In fact, the physical origin of the system decoherence, which follows from the solution of the master equation, is the establishing of quantum entanglement between the system and the bath. Clearly, entanglement prohibits  factorisation of the total density matrix.  
Fortunately,  one can justify a weaker than Eq.~(\ref{3}) assumption, namely,
%*****************************************
\begin{equation}
\label{4}
{\rm Tr}_B[\Lambda R(t)] = {\rm Tr}_B[\Lambda \rho_B(t)] \rho_S(t) \;,
\end{equation}
where $\Lambda$ is an arbitrary operator defined in the bath Hilbert space \cite{25}. It is easy to see that Eq.~(\ref{4}) formally follows from  Eq.~(\ref{3}), but the opposite is not true. Nevertheless, Eq.~(\ref{4})  together with the assumptions (i) and (ii) suffices to derive the master equation (\ref{2}).

{\em 2.} In the present work we revisit the three assumptions listed above [with  Eq.~(\ref{3}) substituted by  Eq.~(\ref{4})]  by considering a simple microscopic `system$+$bath' model of a two-level system, $H_S=\delta \hat{\sigma}_z$, which is attached to the first site of the Bose-Hubbard chain,
%***************************************
\begin{equation}
\label{5}
H_B=  -\frac{J}{2} \sum_{l=1}^{L-1} \left( \hat{a}^\dag_{l+1}\hat{a}_l +h.c.\right)
  +\frac{U}{2}\sum_{l=1}^L \hat{n}_l(\hat{n}_l-1) \;.
\end{equation}
The latter system is known to be generally chaotic in the sense of Quantum Chaos \cite{Gian91,Stoe99,Haak10} that is reflected in the Wigner-Dyson spectrum statistics and ergodic properties of the eigenstates \cite{66,103}.  As the coupling operator we choose
%***************************************
\begin{equation}
\label{6}
H_{int}=\epsilon( \hat{a}_1^\dagger\hat{a}_2 \hat{\sigma}_{+} + h.c.) \;.
\end{equation}
Thus, the `spin'  flips up if a Bose particle tunnels into the first site of the chain and flips down if the particle tunnels out of this site.  It is shown below that strong ergodic properties of the bath indeed justify the Born-Markov approximation.

\begin{figure}
\includegraphics[width=8.5cm,clip]{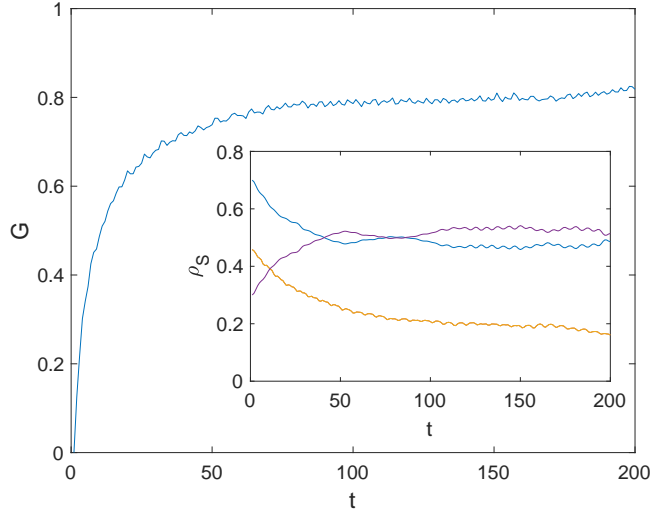}
\caption{Main panel: The quantity (\ref{8}) as the function of time. Inset: the matrix elements of the reduced density matrix $\rho_S(t)$ by the absolute value. The system parameters are $\delta=0.5$ and $\epsilon=0.2$.  The bath parameters are $J=1$, $U=0.8$, $L=7$, and $N=6$. The initial bath state is given by the eigenstate of the Bose-Hubbard Hamiltonian with the energy $E=2.8361$.}
\label{fig1a}
\end{figure}

\begin{figure}
\includegraphics[width=8.5cm,clip]{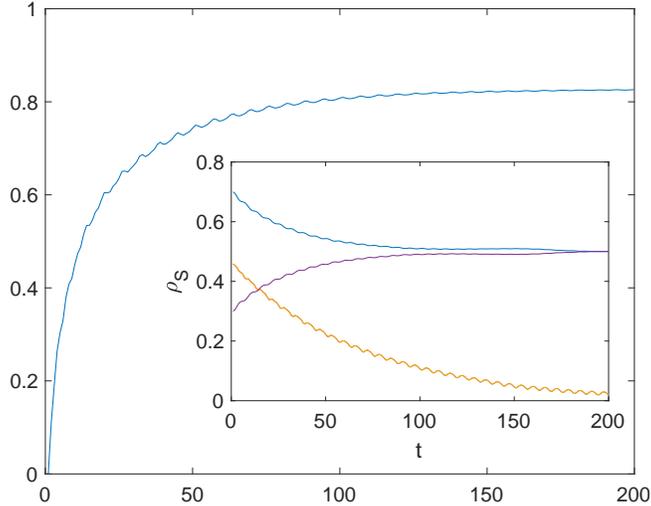}
\caption{The same as in Fig.~\ref{fig1a} yet for the initial bath state $\rho_B(t=0) \sim \sum_j |\Psi_j\rangle \langle \Psi_j |$ where the sum is taken over one hundred eigenstates falling in the energy interval $2.45\le E \le 3.21$.}
\label{fig1b}
\end{figure}

\begin{figure}
\includegraphics[width=8.5cm,clip]{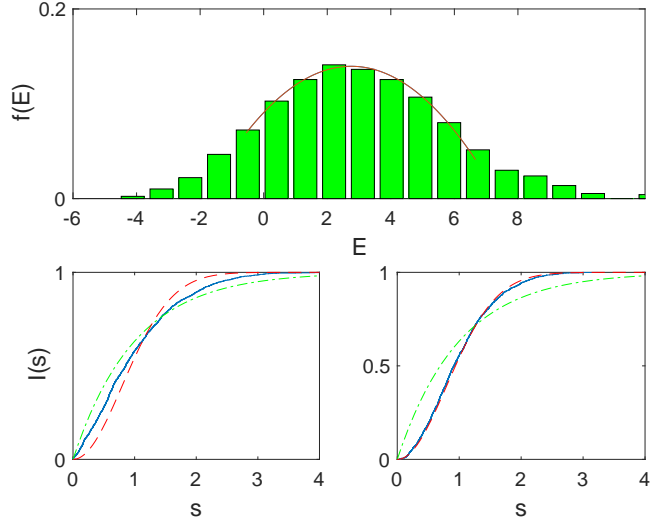}
\caption{Lower panels: Integrated level-spacing distribution for the (unfolded) energy spectrum of the total Hamiltonian $H$ for $\epsilon=0$ (left) and $\epsilon=0.2$ (right). Notice that small $\epsilon$ does not affect the density of state shown in the upper panel.}
\label{fig3}
\end{figure}

{\em 3.} First we illustrate invalidity of Eq.~(\ref{3}). We simulate the dynamics of the total system,
%***************************************
\begin{equation}
\label{7}
R(t)=W(t) R(0) W^\dagger(t) \;,\quad 
%\end{equation}
%\begin{displaymath}
W(t)=\exp(-i H t) \;,\quad H=H_S + H_B + H_{int} \;,
%\end{displaymath}
\end{equation}
for the initial condition given by a product state, i.e., $R(t=0)=\rho_B \otimes \rho_S$. Although the uncorrelated initial state is usually considered to be a rather important  assumption for validity of the master equation, here we use it exclusively to demonstrate the onset of quantum entanglement. To be certain we choose $\rho_S=|\psi\rangle \langle \psi |$ where $|\psi\rangle$ is given by a coherent superposition of the two system eigenstates,  $|\psi\rangle =\sqrt{0.7} |\uparrow\rangle+\sqrt{0.3} |\downarrow\rangle$, and $\rho_B=|\Psi\rangle \langle \Psi |$ where $|\Psi\rangle$ is an eigenstate of the Bose-Hubbard Hamiltonian  with the energy $E=\langle \Psi |H_B|\Psi\rangle$ from the centre of its energy spectrum [see Fig.~\ref{fig3}(a) below]. In this case the Bose-Hubbard chain acts as a high-temperature bath, inducing relaxation of the reduced density matrix $\rho_S(t)$ into a diagonal matrix, see inset in Fig.~\ref{fig1a}.  The decay of the off-diagonal elements reflects the onset of entanglement between the system and the bath which we characterise by the quantity
%***************************************
\begin{equation}
\label{8}
G(t)=\frac{|R(t)-\rho_B(t) \otimes \rho_S(t)|}{ |R(t)|} \;,
\end{equation}
where the modulus sign denotes the sum of all matrix elements taken by the absolute value. (Another measure of entanglement based on the information entropy is discussed in Appendix A.) A rapid growth of entanglement is clearly seen in Fig.~\ref{fig1a}.  We mention that time-fluctuations of the matrix elements of $\rho_S(t)$ can be greatly reduced by choosing  a few eigenstates in a narrow energy interval as the initial bath state, i.e.,  $\rho_B(t=0) \sim \sum_j |\Psi_j\rangle \langle \Psi_j |$.  This additional average also reveals the exponential law for the decay  of the off-diagonal elements with the decay rate proportional to $\epsilon^2$, see   Fig.~\ref{fig1b}.
%###################################### fig1
%###################################### fig1

A comment on the value of the coupling constant $\epsilon$ is in turn. As it was mentioned in the introductory part of the paper, $\epsilon$ has to be small to justify the Markovian master equation (\ref{2}). However, since we work with a finite bath, it has not to be smaller than some critical value $\epsilon_{cr}$.  One finds this critical value by analysing the spectrum statistics of the total Hamiltonian.  The lower-left panel in Fig.~\ref{fig3} shows the integrated level-spacing distribution for the parameters of Fig.~\ref{fig1a} yet $\epsilon=0$. The calculated   distribution is given the direct sum of two independent GOE spectra. When $\epsilon$ is increased from zero to $\epsilon_{cr}$ the level-spacing distribution converges to that for a single GOE, see the lower-right panel in Fig.~\ref{fig3}. Remarkably,  the critical $\epsilon$ decreases with an increase  of the bath size.  For example, $\epsilon_{cr}=0.2$ for $(N,L)=(5,6)$ (dimension of the bath Hilbert space ${\cal N}_B=504$),   $\epsilon_{cr}=0.1$ for $(N,L)=(6,7)$ (${\cal N}_B=924$), and $\epsilon_{cr}\le0.05$ for   $(N,L)=(7,8)$ (${\cal N}_B=3432$). Thus, the existence of the lower boundary for the coupling constant is a finite-size effect.
%###################################### fig3

\begin{figure}
\includegraphics[width=8.5cm,clip]{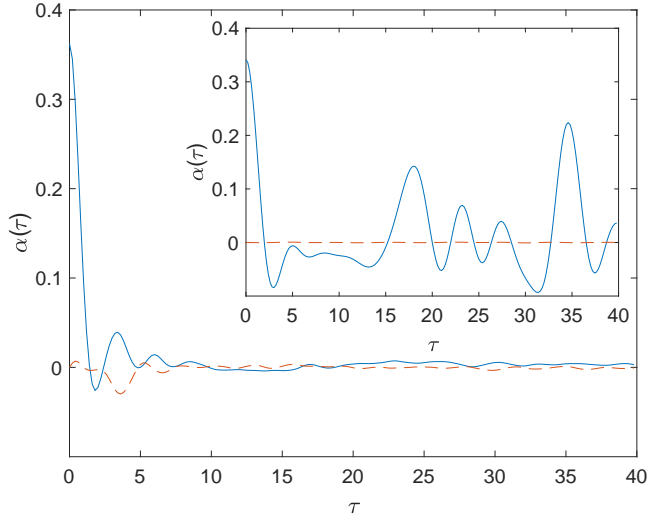}
\caption{Main panel: The real (solid line) and imaginary (dashed line) parts of the correlation function (\ref{9}) as the function of $\tau$ at $t=100$. Parameters are the same as in Fig.~\ref{fig1b}. Inset: Correlation function of the regular bath which one obtains by setting $U=0$ in the Bose-Hubbard Hamiltonian.}
\label{fig4}
\end{figure}

{\em 4.} Next we discuss the correlation time of the bath. The relevant to the master equation correlation function of the bath has the form
%***************************************
\begin{equation}
\label{9}
\alpha(\tau,t)={\rm Tr}_B[\Lambda(\tau)\rho_B(t)] \;, \quad
%\end{equation}
%\begin{displaymath}
\Lambda(\tau)= \exp(-iH_B\tau)\hat{a}_1^\dag\hat{a}_2 \exp(iH_B\tau)\hat{a}_2^\dag\hat{a}_1 \;.
%\end{displaymath}
\end{equation}
The function $\alpha(\tau,t)$ is shown in Fig.~\ref{fig4} where we fixed time to $t=100$. It is seen in Fig.~\ref{fig4} that correlations decay within the characteristic time $\tau^*\approx4$ which is short enough comparing to the characteristic time-scale of the system dynamics shown in Fig.~\ref{fig1b}.

We stress that the decay of correlations is entirely due to chaotic nature of the bath. For the sake of comparison the inset in Fig.~\ref{fig4} shows the correlation function for $U=0$ where the Bose-Hubbard model reduces to $L$ non-interacting linear oscillators. We also mention that for chosen $t=100$ the system and the bath are already entangled so that  $\rho_B(t=100)$ strongly differs from $\rho_B(t=0)$ if compared matrix element against matrix element. Yet, we get the same result if we choose $t=0$.  This clarifies the meaning of the statement that one can neglect back action of the system on the bath -- it does not imply that $\rho_B(t)\approx \rho_B(0)$ but that the ergodic properties of the bath remain unchanged.  A more detailed analysis of the correlation function based on the statistics of the transition matrix elements of the interaction Hamiltonian $H_{int}$ is given in Appendix B.
%###################################### fig1

\begin{figure}
\includegraphics[width=8.5cm,clip]{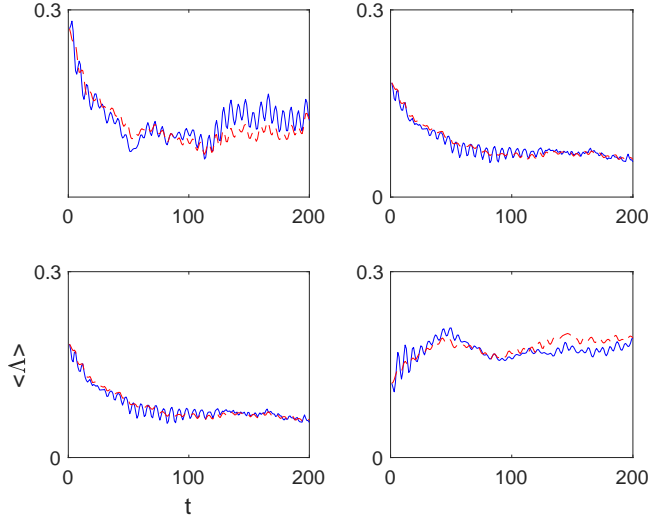}
\caption{Four quantities calculated  according to the left-hand-side (solid lines) and the right-hand-side (dashed lines) of Eq.~(\ref{4}).}
\label{fig2}
\end{figure}

{\em 5.} Finally we check the validity of Eq.~(\ref{4}). As the operator $\Lambda$ in this equation we consider $\Lambda(\tau=0)=\hat{a}_1^\dagger\hat{a}_2 \hat{a}_2^\dagger\hat{a}_1$. The solid and dashed lines in four panels in Fig.~\ref{fig2} show four quantities calculated  according to the left-hand-side and the right-hand-side of Eq.~(\ref{4}). A reasonable agreement is noticed. It should be stressed that the discussed equation is not mathematically exact and may hold only  with some accuracy. Originally it was deduced in Ref.~\cite{25} by appealing to the mixing property of classical chaotic systems and the quantum-classical correspondence. The results in Fig.~\ref{fig2} indicate that this equation also holds in the case where the quantum system has no obvious classical counterpart.  
%###################################### fig2

{\em 6.} To summarise, we considered the microscopic model of the bath given by the Bose-Hubbard model in the parameter region where it is chaotic in the sense of Quantum Chaos. Unlike the other popular microscopic model of the bath -- an infinite number of linear oscillators -- the chaotic Bose-Hubbard bath has strong ergodic properties that lead to wide-scale quantum entanglement not only between the bath modes but also between the bath modes and the system of interest which is coupled to the bath.  Importantly, these ergodic properties justify all assumptions which one needs to derive the Markovian master equation (\ref{2}) for the reduced density matrix of the system.  

The author acknowledge discussions with D. N. Maksimov and financial support of Russian Science Foundation (RU) through the grant N19-12-00167.

%%%%%%%%%%%%%%%%%%%%%%%%%%%%%%%%%%%%%%%%%%%

\newpage
%%%%%%%%%%%%%%%%%%%%%%%%%%%%%%%%%%%%%%%%%%%%%%%
%\section{Appendix A}
%\vspace*{0.5cm}
{\bf Appendix A} 
\vspace*{0.5cm}

\begin{figure}[b]
\includegraphics[width=8.5cm,clip]{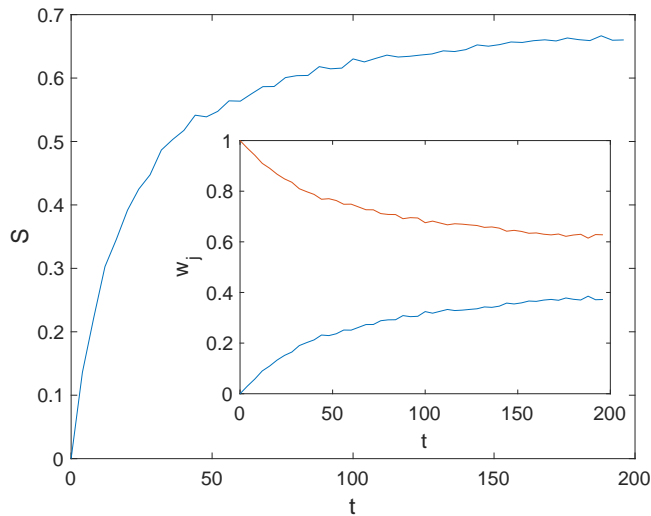}
\caption{Von Neumann entropy, main panel, and eigenvalues of $\rho_S(t)$, inset, as the function of time for the parameters of Fig.~\ref{fig1a}. }
\label{fig5}
\end{figure}

For the initial condition used in  Fig.~\ref{fig1a} the total density matrix $R(t)$ is a pure state for any time. However, the reduced density matrices $\rho_S(t)$ and $\rho_B(t)$ becomes mixed states in course of time. Using the spectral decomposition we have
%***************************************
\begin{equation}
\label{a0}
\rho_S(t)= \sum_{n=1}^2 w_n(t) |\phi_n(t)\rangle \langle \phi_n(t) | \;,
\end{equation}
where $w_n(t)$ and  $|\phi_n(t)\rangle$ are eigenvalues and eigenstates of the $2\times 2$ matrix $\rho_S(t)$. The commonly accepted  quantitative characteristic of the entanglement is the information entropy $S(t)=-\sum_n w_n \log(w_n)$.  Dynamics of the entropy $S(t)$ and eigenvalues $w_n(t)$ are exemplified in Fig.~\ref{fig5}.
%###################################### fig2
%###################################### fig2
\begin{figure}
\includegraphics[width=8.5cm,clip]{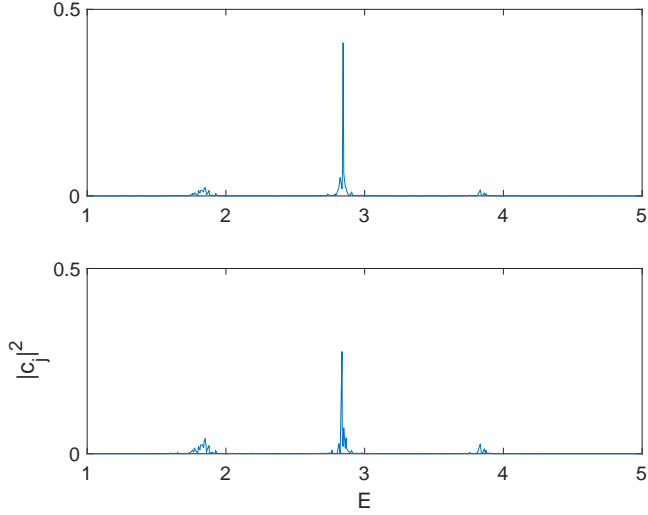}
\caption{ Expansion coefficients $c_j^{(n)}(t)$ at $t=200$ of the two states  $|\Psi_n \rangle$ which appear in the spectral decomposition of the reduced matrix $\rho_B(t)$.}
\label{fig6}
\end{figure} 

Analogously, for the reduced density matrix of the bath we have
%***************************************
\begin{equation}
\label{a1}
\rho_B(t)= \sum_{n=1}^{{\cal N}_B} w_n(t) |\Phi_n(t)\rangle \langle \Phi_n(t) | \;,
\end{equation}
where the summation formally runs to ${\cal N}_B$ -- the dimension of the Hilbert space of the Bose-Hubbard Hamiltonian. However, according to the Schmidt theorem for the pure $R(t)$ only two eigenvalues $w_n(t)$ differ from zero and they coincide with the eigenvalues of $\rho_S(t)$. Thus, the information entropy of the bath is the same. The difference appears in the structure of the eigenstates. Fig.~\ref{fig6} shows the expansion coefficients of the states   $|\Phi_1(t)\rangle$ and   $|\Phi_2(t)\rangle$ in the bath energy basis,
%***************************************
\begin{equation}
\label{a2}
|\Phi_n(t)\rangle= \sum_{j=1}^{{\cal N}_B} c_j^{(n)}(t) |\Psi_j\rangle  \;.
\end{equation}
It is seen that the system-bath interaction admixes to the initial bath state $|\Psi_E\rangle$ with the energy $E=2.8361$ the other eigenstates of $H_B$  which form three groups separated by the energy interval $2\delta$. Since $\delta$ is assumed to be small comparing to the width of the bath spectrum, all these states have the same ergodic properties as the initial state. This implies, in particular, that the mean value of any observable, $\langle  \Lambda \rangle = \langle \Psi_ j \Lambda |\Psi_j\rangle |$, is the same up to statistical fluctuations. Thus,
%***************************************
\begin{equation}
\label{a4}
{\rm Tr}[\Lambda \rho_B(t)]= {\rm Tr}[\Lambda \rho_B(t=0)]  \;.
\end{equation}
This equation is a formalisation of the statement about the negligible effect of the system on the bath.

%%%%%%%%%%%%%%%%%%%%%%%%%%%%%%%%%%%%%%%%%%%%%%%
%\section{Appendix B}
\vspace*{0.5cm}
{\bf Appendix B} 
\vspace*{0.5cm}

The particular shape of the correlation function in Fig.~\ref{4} is determined by the statistics of the transition matrix elements of the interaction Hamiltonian. In fact, in terms of the eigenstates and eigenenergies of the Hamiltonian $H_B$ Eq.~(\ref{9}) can be written as
%***************************************
\begin{equation}
\label{b1}
\alpha(\tau)=\sum_k |\langle \Psi_k|\hat{a}_1^\dag\hat{a}_2|\Psi_j\rangle|^2 e^{i(E_k-E_j)\tau} \;,
\end{equation}
where we set $t=0$ and $|\Psi_j\rangle$ is the initial bath state. Labelling the states $| \Psi_k \rangle$ by their energies instead of the index $k$,  Eq.~(\ref{b1}) takes the form
%***************************************
\begin{equation}
\label{b2}
\alpha(\tau)=\int {\rm d}E  V(E,E_j) e^{i(E-E_j)\tau} \;,
\end{equation}
where
%***************************************
\begin{equation}
\label{b3}
V(E,E_j)=\overline{|\langle \Psi_k|\hat{a}_1^\dag\hat{a}_2|\Psi_j\rangle|^2}
\end{equation}
and the bar denotes the average over a small energy interval ${\rm d}E$. Notice that the function (\ref{b3}) depends only on the energy difference $\Delta E=E-E_j$. Thus, we have  additional average over $E_j$ if the initial bath state if a mixed state.
%###################################### fig2
\begin{figure}
\includegraphics[width=8.5cm,clip]{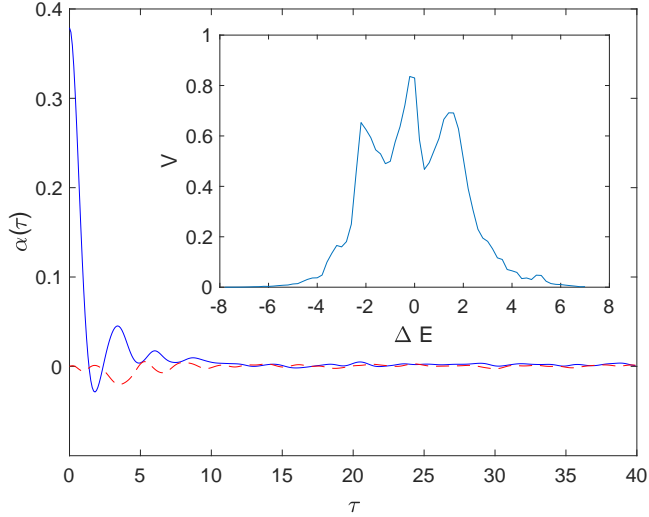}
\caption{Main panel: The correlation function of the bath calculated according to Eq.~(\ref{b1}). Inset: Distribution function for the transition matrix elements. To reduce statistical fluctuations here we consider a larger bath $(N,L)=(7,8)$ where the dimension of the bath Hilbert space is ${\cal N}_B=3432$.}
\label{fig7}
\end{figure}

The function $V=V(\Delta E)$ is depicted in the inset in Fig.~\ref{7}. It is a structured Gaussian where the number and positions of local peaks are determined by the value of the interaction constant $U$. This fine structure of $V(\Delta E)$ is responsible for oscillations of $\alpha(\tau)$ whereas the Gaussian envelope determines the overall decay of the correlation function as
%***************************************
\begin{equation}
\label{b4}
\alpha(\tau) \sim \exp\left[-\left(\frac{\tau}{\tau^*}\right)^2\right] \;,
\end{equation}
where the correlation time $\tau^*$ is inverse proportional to the Gaussian width.

\end{document}